\documentclass[reference]{aa}
\usepackage{graphicx}
\usepackage{xcolor}
\bibpunct{(}{)}{;}{a}{}{,} 

\begin{document}

\title{High-energy  emitting BL Lacs and high-energy neutrinos.}
\subtitle{Prospects for the direct association with IceCube and KM3NeT}


\author{C. Righi \inst{\ref{inst1} \and \ref{inst2}}
            \and F. Tavecchio \inst{\ref{inst2}}
            \and D. Guetta\inst{\ref{inst3} \and  \ref{inst4}}}

\institute {
Dipartimento di Fisica, Universit\`a dell'Insubria --  Via Valleggio, 11 - 22100 Como, Italy \label{inst1}
\and 
INAF--Osservatorio Astronomico di Brera, Via E. Bianchi 46, I--23807 Merate, Italy \label{inst2}
\and 
INAF--Osservatorio Astronomico di Roma, via Frascati 33, I--00040 Monte Porzio Catone, Italy \label{inst3}
\and 
Department of Physics Optical Engineering, ORT Braude, P.O. Box 78, Carmiel, Israel \label{inst4}
}

 \date{ 25 July 2016 / 25 September 2016}


\abstract
{The origin of the high-energy flux of neutrinos detected by IceCube is still unknown. Recent works report the evidence for a possible positional correlation between the reconstructed neutrino arrival directions and the positions in the sky of low power, high-energy emitting BL Lac objects (HBL).}
{Assuming that $\gamma$-ray emitting HBL form the bulk of the sources of high-energy neutrinos above 100 TeV, we intend to calculate the number of events expected to be detected for each source by IceCube and KM3NeT.}
{Based on a simple theoretically-motivated framework inspired by the structured jet scenario for these sources, we postulate a direct proportionality between high-energy $\gamma$-ray and neutrino fluxes. We calculate the expected neutrino event rate for the HBL sources of the Second Fermi-LAT Catalog of High-Energy Sources (2FHL) for IceCube and the presently under construction KM3NeT using declination-dependent and exposure-weighted effective areas.}
{We provide a list of 2FHL HBL with the calculated number of events. For IceCube,  the derived count rate for several sources is relatively high, of the order of $\lesssim$1 yr$^{-1}$, consistently with the recent findings of a possible positional correlation. For KM3NeT the calculated rates are higher, with several sources with expected rate exceeding 1 yr$^{-1}$. This, coupled with the improved angular resolution, implies that the HBL origin can be effectively tested with few years of observation of KM3NeT (and IceCube Gen2, for which similar performances are foreseen) through the direct association of neutrinos and single HBL.}
{Our results show that if -- as hinted by recent works -- HBL represent a possible population of high-energy neutrino emitters, several single sources should be identified in few years of exposure of KM3NeT, highlighting the importance of the improved angular resolution anticipated for KM3NeT and IceCube Gen2.}



\maketitle
\section{Introduction}

The cosmic sources responsible for the extraterrestrial neutrino flux detected by IceCube at PeV energies (Aartsen et al. 2013, 2014, 2015a) are still unknown.  The substantial isotropy of the flux (with only a non significant small excess in the direction of the galactic center) is consistent with an extragalactic origin, although a slight north-south  intensity and hardness asymmetry could  hint to a possible contribution from a soft galactic component, superseded by a harder extragalactic component emission above $\approx 100$ TeV  (e.g. Ahlers \& Murase 2014, Neronov \& Semikoz 2015, Aartsen et al. 2015, Palladino \& Vissani 2016).  Among the possible extragalactic astrophysical sources there are propagating comic rays (e.g., Essey et al. 2010, Kalashev et al. 2013), star-forming and starburst galaxies (e.g., Loeb \& Waxman 2006, Wang, Zhao, \& Li 2014, Tamborra et al. 2014), galaxy clusters (e.g., Murase \& Beacom 2013, Zandanel et al. 2015), $\gamma$--ray burst (e.g., Waxman \& Bahcall 1997, Petropoulou et al. 2014) and active galactic nuclei (AGN, e.g., Mannheim 1995, Atoyan \& Dermer 2003, Kimura et al. 2014, Kalashev et al. 2014, Petropoulou et al. 2015, 2016).

Among AGN, blazars (e.g. Urry \& Padovani 1995) are often considered the most probable candidates. Blazars (further divided in flat spectrum radio quasars, FSRQ, and BL Lac objects) are AGN presenting two jets ejected at relativistic speeds in opposite directions, one of which is well aligned with the line of sight to the Earth. In this geometry, relativistic effects greatly enhance the observed intensity (relativistic beaming), making these sources among the brightest extragalactic sources. Because of the beaming, the emission observed from blazars (extending over the entire electromagnetic spectrum with a characteristic double-humped shape when plotted in the $\nu F_{\nu }$ representation - the so-called spectral energy distribution, SED) is dominated by the non-thermal continuum produced in the jet. Leptonic models attribute the entire emission to relativistic electrons/pairs radiating through synchrotron and inverse Compton mechanisms, responsible for the low and the high energy SED peaks, respectively (e.g. Ghisellini et al. 1998). In the hadronic scenario, instead, the high-energy peak is linked to high-energy hadrons co-accelerated with electrons, cooling through the synchrotron or the photo-meson channel (e.g. B{\"o}ttcher et al. 2013). Blazars jets appear ideal sites to accelerate hadrons (protons, for simplicity) to the energy $E_{\rm p}\approx 10^{17}$ eV required to produce PeV neutrinos, most likely via the photomeson reaction ($p+\gamma \to X +\pi$),  followed by the prompt decay of the charged pions ($\pi ^{\pm}\to \mu^{\pm}+\nu_{\mu}\to e^{\pm} + 2\nu_{\mu} +\nu_{\rm e}$; hereafter we do not distinguish among $\nu$ and $\bar{\nu}$). In fact the possible role of blazars has been recently highlighted by the results of Kadler et al. (2016) and Padovani et al. (2016). Kadler et al. (2016) report a tempting correlation between the arrival time of one of the neutrino with the highest reconstructed energy ($\sim 2$ PeV) and an exceptional outburst  phase of the FSRQ PKS B1414-418, which lies in the (large, radius $\sim 16^{\circ}$) IceCube uncertainty region for this event. Padovani et al. (2016) on the other hand, improving a previous work by Padovani \& Resconi (2014), have presented the evidence for a significant (random expectation level of $\approx 0.4\%$) spatial correlation between the reconstructed arrival direction of neutrinos (including both hemispheres) and BL Lac objects emitting very high-energy $\gamma$ rays ($>50$ GeV). No correlation is instead found with other classes of blazars, such as FSRQ or BL Lacs with larger luminosity. Taken together, these two results are quite intriguing and puzzling, since powerful FSRQ and high-energy emitting BL Lacs (hereafter HBL, standing for highly peaked BL Lac objects), are objects characterized by rather different physical properties that  lie at the opposite sides of the so-called blazar sequence (Fossati et al. 1998), relating the spectral properties of the emission of blazars with their luminosity. 

At a first sight, the FSRQ environment seems to offer the best conditions (high jet power, dense target radiation fields) to account for neutrino production through the photomeson channel ($pp$ reactions are unlikely in the low-energy jet environment), while BL Lac seem disfavored, mainly because their low luminosity hints to inefficient photomeson production (e.g. Murase et al. 2014). However, in a previous paper (Tavecchio, Ghisellini \& Guetta 2014, hereafter Paper I), we showed that, if the jet is characterized by a velocity structure, i.e. the flow is composed by a fast spine surrounded by a slower sheath (or layer), the neutrino output from HBL can be highly boosted with respect to the one-zone models and the cumulative emission of the HBL population could match the observed intensity with an acceptable value of the cosmic ray power for the jet. The existence of a velocity structure of the jet has been previously considered as a possible solution for several issues related to TeV emitting BL Lacs and to unify the BL Lacs and radiogalaxy populations  (e.g. Chiaberge et al. 2000, Meyer et al. 2011, Sbarrato et al. 2014).  Direct  radio VLBI imaging of jets both in low-power radiogalaxies (e.g. Nagai et al. 2014, M{\"u}ller et al. 2014) and BL Lac  (e.g., Giroletti et al. 2004, Piner \& Edwards 2014), often showing a ``limb brightening"  transverse structure, provides a convincing observational support to this idea, also corroborated by numerical simulations (e.g. McKinney 2006, Rossi et al. 2008).  The increased neutrino (and inverse Compton $\gamma$-ray) production efficiency in the spine--layer structure is based on the fact that for particles flowing in the faster region the radiation field produced in the layer is amplified by the relative motion between the two structures (e.g. Ghisellini et al. 2005, Tavecchio \& Ghisellini 2008). In this condition, the density of the soft photons in the spine rest frame -- determining the proton energy loss rate and hence the neutrino luminosity -- can easily exceed that of the radiation produced locally, the only radiative component considered in the one--zone modeling of BL Lacs (for FSRQ, instead, the photon field is thought to be dominated by the radiation coming from the external environment). In Tavecchio \& Ghisellini (2015) we relaxed the condition that only HBL jets are able develop an important layer, assuming  that all BL Lacs jets are characterized by a structure region and that the layer radiative luminosity and the cosmic ray power are both proportional to the jet power. 

In searching for a {\it direct} association between neutrinos and possible sources one can exploit the temporal coincidence between the neutrino detection and  high-state/flares of a source (e.g., Kadler et al. 2016, Halzen \& Kheirandish 2016) and/or the coincidence between the reconstructed arrival direction of  neutrinos and the position of a putative source in the sky (e.g. Padovani et al. 2016). The latter works best when applied to the events detected through up-going muons, which provide the best angular resolution. The practical application of this methods is generally based on the use of a pre-selected list of possible neutrino source candidates (e.g. Adrian-Martinez et al. 2016a). Along these lines, in this paper, motivated by the recent results by Padovani et al. (2016), we aim at reconsidering the possible production of neutrinos by HBL, focusing in particular on the  fluxes expected for present (IceCube) and future (KM3NeT) neutrinos observatories. Assuming a simple phenomenological framework inspired by the spine-layer scenario and supported by the Padovani et al. findings (\S2), we connect the putative neutrinos fluxes to the observed high-energy gamma-ray fluxes (\S3) and then we provide the expected neutrinos counts (\S4).

Throughout the paper, the following cosmological  parameters are assumed: $H_0=70$ km s$^{-1}$ Mpc$^{-1}$, $\Omega_{\rm M}=0.3$, $\Omega_{\Lambda}=0.7$.  We  use the notation $Q=Q_X \, 10^X $ in cgs units.
    
\section{Theoretical framework}

\begin{figure}
  \resizebox{\hsize}{!}{\includegraphics{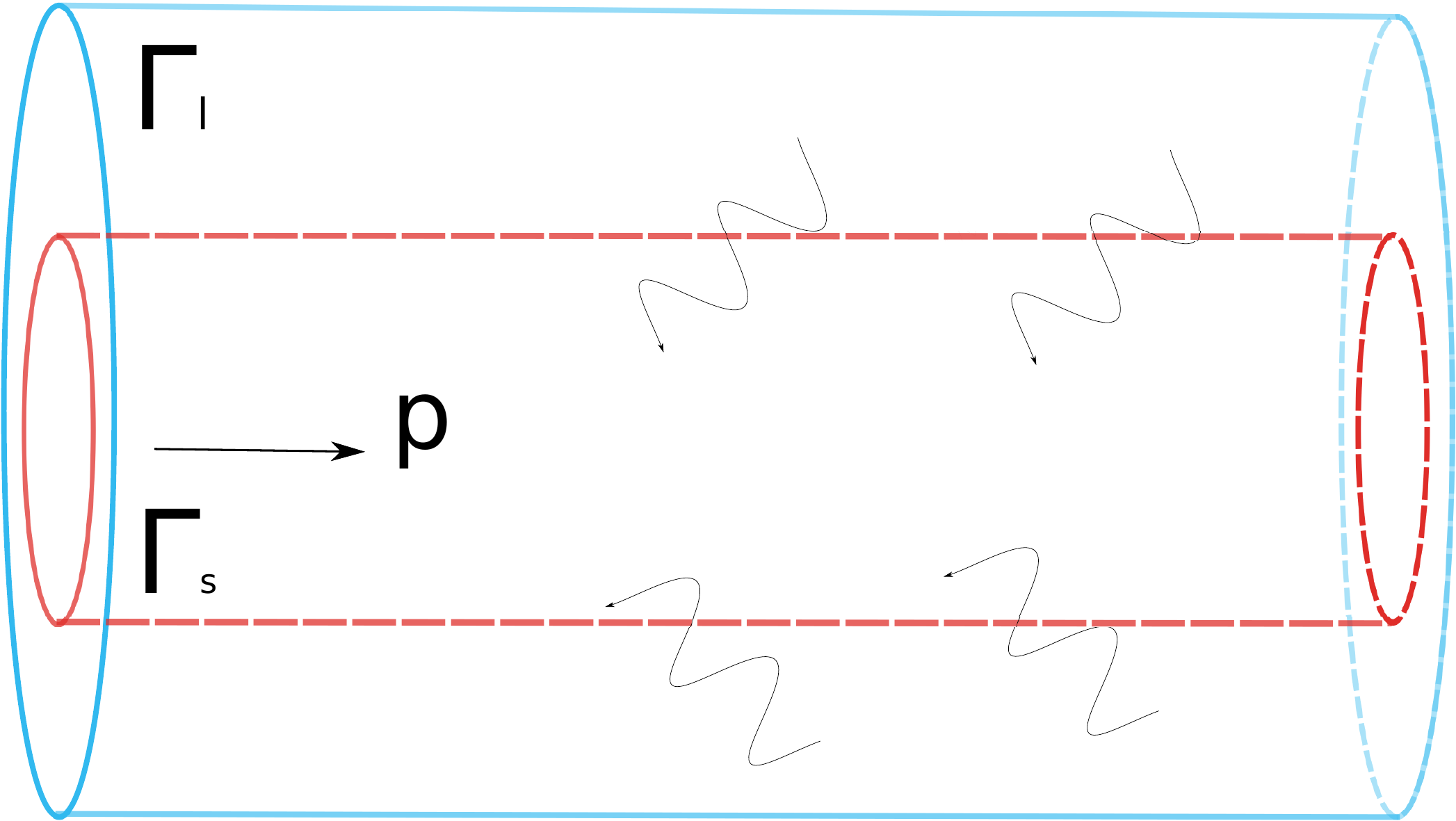}}
  \caption{Sketch of spine-layer geometry. A faster inner core, or spine, is surrounded by a low-velocity layer. Both regions emit low-energy synchrotron photons. Due to the relative motion, the low-energy emission of the layer is amplified in the spine frame and dominate the photo-meson cooling of high-energy protons.}
  \label{fig:esempio}
\end{figure}

We refer to Ghisellini et al. (2005) and Paper I for a complete description of the spine-layer jet scenario and its application to the neutrino production. Here we just recall that the jet geometry (see Fig.\ref{fig:esempio}) is approximated by two concentric cylinders with different bulk Lorenz factors $\Gamma_s > \Gamma_l$. If we observe the jet at an angle of view $\theta_v$, the two relativistic Doppler factor $\delta$ determining the amplification of the emission are defined as $\delta_{s,l}=[\Gamma_{s,l}(1-\beta_{s,l} \cos\theta_v)]^{-1}$, in which $\beta=v/c$ . In the following, primed symbols indicate quantities measured in the spine reference frame. 

We adopt a leptonic scenario for the blazar {\it electromagnetic} emission, in which the observed radiation (dominated by the highly boosted spine emission) is entirely attributed to leptons directly accelerated in the jet. This is equivalent to assume that any electromagnetic component associated to hadronic reactions (and hence to neutrino emission) do not dominate the SED (note that the high-energy $\gamma$ rays from the $\pi^0$ decay are promptly absorbed through scattering with the soft photons and, after reprocessing, leave the jet at much lower energies, in the MeV-GeV band). The high-energy emission is dominated by the boosted soft photons produced in the layer inverse-Compton scattered to $\gamma$-ray energies by the relativistic electrons flowing into the spine (also responsible for the low-energy SED synchrotron bump). In this sense, the electromagnetic and neutrino outputs derive from two different (but not independent, see below) channels.

Consider now the emitted neutrino luminosity. As detailed in Paper I, the (observed) neutrino luminosity (single flavor, assuming a $\nu_e:\nu_{\mu}:\nu_{\tau} = 1:1:1$  flavor composition at the Earth) at the observed energy $E_{\nu}$ can be well approximated by:
\begin{equation}
E_\nu L_\nu(E_\nu)\simeq \frac{1}{8}f_{p\gamma}(E^{\prime}_{\rm p})\,  E^{\prime}_{\rm p}Q_{\rm p}^{\prime}(E^{\prime}_{\rm p}) \, \delta^4_s
\end{equation}
where $E^{\prime}_{\rm p}Q_{\rm p}^{\prime}(E^{\prime}_{\rm p})$ is the power injected into protons of energy $E_{\rm p}^{\prime}$ and $E^{\prime}_{\nu}=E_{\nu}/\delta_s\simeq E^{\prime}_{\rm p}/20$. The factor $f_{p\gamma}(E^{\prime}_{\rm p})$ measures the efficiency of the photomeson losses and it is defined as the ratio of the dynamical timescale to the photomeson cooling time:
\begin{equation}
f_{p\gamma}(E^{\prime}_{\rm p})=\frac{t^{\prime}_{\rm dyn}}{t^{\prime}_{p\gamma}(E^{\prime}_{\rm p})}
\end{equation}
where:
$t^{\prime}_{p\gamma}(E^{\prime}_{\rm p})=[c \, \langle n^{\prime}_{\rm ph}(\epsilon^{\prime}) \, \sigma _{p\gamma}(\epsilon ^{\prime},E^{\prime}_{\rm p}) K(\epsilon ^{\prime},E^{\prime}_{\rm p})\rangle]^{-1}$, $\sigma _{p\gamma}$ being the cross section, $K$ the inelasticity and $n^{\prime}_{\rm ph}$ is the target photon number density.

The total, energy integrated, neutrino luminosity can be expressed as:
\begin{equation}
L_\nu= \epsilon_{\rm p}Q_{\rm p}^{\prime} \, \delta^4_s,
\label{lp}
\end{equation}
where the total CR injected power is:
\begin{equation}
Q_{\rm p}^{\prime} = \int Q_{\rm p}^{\prime}(E^{\prime}_{\rm p}) \, dE^{\prime}_{\rm p}
\end{equation}
and the averaged efficiency $\epsilon_{\rm p}$ is:
\begin{equation}
\epsilon_{\rm p}= \frac{1}{Q_{\rm p}^{\prime}}\int f_{p\gamma}(E^{\prime}_{\rm p})\, Q_{\rm p}^{\prime}(E^{\prime}_{\rm p}) \, dE^{\prime}_{\rm p}
\end{equation}
  
In general, the IC (i.e. high-energy $\gamma$-ray) luminosity can be formally expressed in exactly the same way: 
\begin{equation}
L_{\gamma}= \epsilon_{\rm e}Q_{\rm e}^{\prime} \, \delta^4_s,
\label{le}
\end{equation}
where now $\epsilon_{\rm e}$ and $Q_{\rm e}^{\prime}$ refer to the relativistic electrons.
Using Eq. \ref{lp} and \ref{le} one can write  the ratio of the gamma-ray and neutrinos fluxes of a given source as:
\begin{equation}
\label{fluxflux}
\frac{F_{\nu}}{F_{\gamma}}=\frac{L_{\nu}}{L_{\gamma}} = \frac{\epsilon_{\rm p}}{\epsilon_{\rm e}} \, \frac{Q^{\prime}_{\rm p}}{Q^{\prime}_{\rm e}}
\end{equation}

In the spine layer scenario, the soft radiation field in the spine frame is dominated by the relativistically boosted layer radiation. In these conditions, both efficiencies, $\epsilon_{\rm p}$ and $\epsilon_{\rm e}$ depend on the same photon field, $n^{\prime}_{\rm ph,l}$ and thus their ratio, $\epsilon_{\rm p}/\epsilon_{\rm e}\equiv \xi_{\rm ep}$ depends only on the details of the injection and cooling processes. As a zero-order approximation, one can assume that these properties are universal for all the (quite similar) HBL jets, namely that $\xi_{\rm ep}$ is on average constant (with, of course, some dispersion) in the HBL population.
Furthermore we find reasonable to assume that the ratio between the power injected into relativistic electrons and that into high-energy protons is, on average, the same in different sources, both depending 
on the total power carried by the jet, $P_{\rm jet}$, i.e. $Q^{\prime}_{\rm p}=\eta_{\rm p} P_{\rm jet}$ and $Q^{\prime}_{\rm e}=\eta_{\rm e} P_{\rm jet}$, so that $Q^{\prime}_{\rm p}/Q^{\prime}_{\rm e}= \eta_{\rm p}/\eta_{\rm e}\approx const$. With these assumptions, we derive that $F_{\nu}/F_{\gamma}$ is, on average, the same in all HBL, $F_{\nu}/F_{\gamma}=\xi_{\rm ep} \eta_{\rm p}/\eta_{\rm e}\equiv k_{\nu\gamma}$.

Therefore, in our scheme, {\it the bolometric neutrino flux from a given HBL is directly proportional to its high-energy gamma-ray flux}, $F_{\nu}= k_{\nu\gamma} F_{\gamma} $. We remark that this theoretically-inspired assumption in consistent with the results of Padovani et al. (2016), which found that the positional correlation between neutrinos and 2FHL sources holds  at the brightest fluxes.

\section{Calculation}

The Second Catalog of Hard Fermi-LAT Sources (2FHL in short, Ackermann et al. 2016a) includes all the sources detected at energies above $50$ GeV by the Large Area Telescope onboard {\it Fermi} in 80 months of data. The high-energy band covered by the 2FHL  matches well the expected maximum of the IC component produced by the spine. Hence, it is natural -- based on the discussion of the previous section -- to consider the 2FHL flux a good proxy for $F_{\gamma}$. Therefore, using the relation derive above for each source, $F_{\nu i}=k_{\nu\gamma} \, F_{\gamma i}$ it is possible to derive the expected flux of neutrinos.

The constant can be derived under the assumption that the total neutrino diffuse flux measured by IceCube, $F_{\nu, tot}$ is entirely due to the contribution of the high-energy emitting BL Lacs. Since the neutrino flux for each source is directly proportional to the corresponding gamma-ray flux we can write:
\begin{equation}
F_{\nu, tot}\equiv \sum_i {F_\nu}_i = \sum_i k_{\nu\gamma} {F_\gamma}_i = k_{\nu\gamma} \sum_i F_{\gamma i} = k_{\nu\gamma} F_{\gamma ,tot}.
\label{ftot}
\end{equation}
in which we use the fact that $k_{\nu\gamma}$ is (approximately) the same for all sources.
Here $F_{\gamma ,tot}$ is the total high-energy gamma-ray flux from HBL (see below).

The next step is to convert the neutrino energy flux for each source, ${F_\nu}_i$, to the neutrino number flux, $\Phi_i(E_\nu)$, using:
   \begin{equation}
   {F_\nu}_i=\int_{E_1}^{E_2} \Phi_i(E_\nu)E_\nu \, d E_\nu
   \label{fphi}
   \end{equation} 
where 
the interval $[E_1,E_2]$ is the range of neutrino energies.
We assume that each source emits a neutrino spectrum with the same shape of the overall neutrino spectrum reconstructed through the IceCube detections, i.e. a power law distribution (more on this later):
   \begin{equation}
   \Phi_i(E_\nu)=\phi_i \biggl(\frac{E_\nu}{E^\star}\biggr)^{-\Gamma},
   \label{phinu}
   \end{equation}
   where $E^\star$ is the energy of normalization.
 Therefore, from Eqs. \ref{fphi}-\ref{phinu} we can derive the neutrino number flux normalization $\phi_i$ as:
   \begin{equation}
   \phi_i=F_{\nu i} \; {E^\star}^{-\Gamma} \frac{2-\Gamma}{E_2^{2-\Gamma}-E^{2-\Gamma}_1}
   \end{equation}
   
 Finally the number of neutrino $N_\nu$ expected from a given HBL object of the 2FHL catalogue depends on the rate of high energy neutrino $R_\nu$ and the exposure time $T_{\rm exp}$ as follows:
    \begin{equation} 
   N_\nu=R_\nu T_{\rm exp} =T_{\rm exp}\int_{E_1}^{E_2}{A_{\rm eff} (E_\nu)\Phi_i(E_\nu) dE_\nu}
   \label{numbernu}
   \end{equation}
  where $A_{\rm eff}(E_\nu)$ is the effective area of the neutrino detector.

\begin{table}
\caption{Expected 0.1-10 PeV flux (in units of $10^{-8}$ GeV cm$^{-2}$ s$^{-1}$) and detection rate (yr$^{-1}$) of muon neutrino $R_{\nu}$ for the brightest 2FHL BL Lacs with IceCube at declination $60^{\circ}<\delta<90^{\circ}$, $30^{\circ}<\delta<60^{\circ}$,$0^{\circ}<\delta<30^{\circ}$ respectively. The numbers identify sources in the sky map of Fig. \ref{fig:map}.}
\footnotesize
\centering
\begin{tabular}{c|lcc}
\hline

& Name & $F_\nu$ & $R_{\nu}$  \\
\hline

\multicolumn{3}{c}{$60^{\circ}<\delta<90^{\circ}$} \\
\hline
1 & 1ES1959+650 			&   1.38  &       0.27\\
2 & 1ES0502+675 			&   1.14  &       0.22\\
3 & S50716+71 			&   0.44 &       0.08\\ 
4 & 1RXSJ013106.4+61203 	&   0.25  &      0.05\\
5 & 4C+67.04 				&   0.25  &      0.05\\
6 & Mkn180        			&   0.24  &      0.05\\
7 & MS0737.9+7441 		&   0.13  &      0.02 \\ 
8 & RXJ0805.4+7534 		&   0.08  &      0.02  \\
9 & S40954+65 			&   0.07  &      0.01 \\
10 &S41749+70        		&   0.07  &       0.01\\
\hline
\multicolumn{3}{c}{$30^{\circ}<\delta<60^{\circ}$} \\
\hline
11 & Mkn421 				&  8.77   &        4.89	\\ 
12 & Mkn501 				&  3.41   &        1.90  \\ 
13 & PG1218+304 			&  0.92   &       0.52	 \\ 
14 & 3C66A 				&  0.87   &       0.49	 \\ 
15 & 1H1013+498             	&  0.87   &       0.49 \\ 
16 & 1ES0033+595  			&  0.82   &       0.46	\\ 
17 & 1ES2344+514 			&  0.69   &       0.39 \\ 
18 & 1ES1215+303 			&  0.52   &       0.29 \\ 
19 & B32247+381 			&  0.37   &       0.21	 \\ 
20 & B30133+388 			&  0.35   &       0.19 \\
\hline
\multicolumn{3}{c}{$0^{\circ}<\delta<30^{\circ}$} \\
\hline
21 & PG1553+113 			&   1.89  &        2.47     \\ 
22 & PKS1424+240 			&   1.00  &       1.30     \\   
23 & PG1218+304 			&    0.92 &       1.20     \\   
24 & TXS0518+211 			&    0.87 &       1.14      \\  
25 & 1ES0647+250 			&    0.75 &       0.99  	 \\ 
26 & 1ES1215+303 			&    0.52 &       0.69	  \\  
27 & RXJ0648.7+1516 		&    0.45 &       0.59	 \\  
28 & 1RXSJ194246.3+10333 	&    0.41 &       0.54     \\  
29 & RBS0413 				&    0.32 &       0.42    \\  
30 & 1H1720+117 			&    0.25 &       0.33     \\  
\hline

\end{tabular}
\label{TableIce}
\end{table}

\begin{figure}
\hspace{-0.1truecm}
 \includegraphics[width=9.3cm,height=8.5cm]{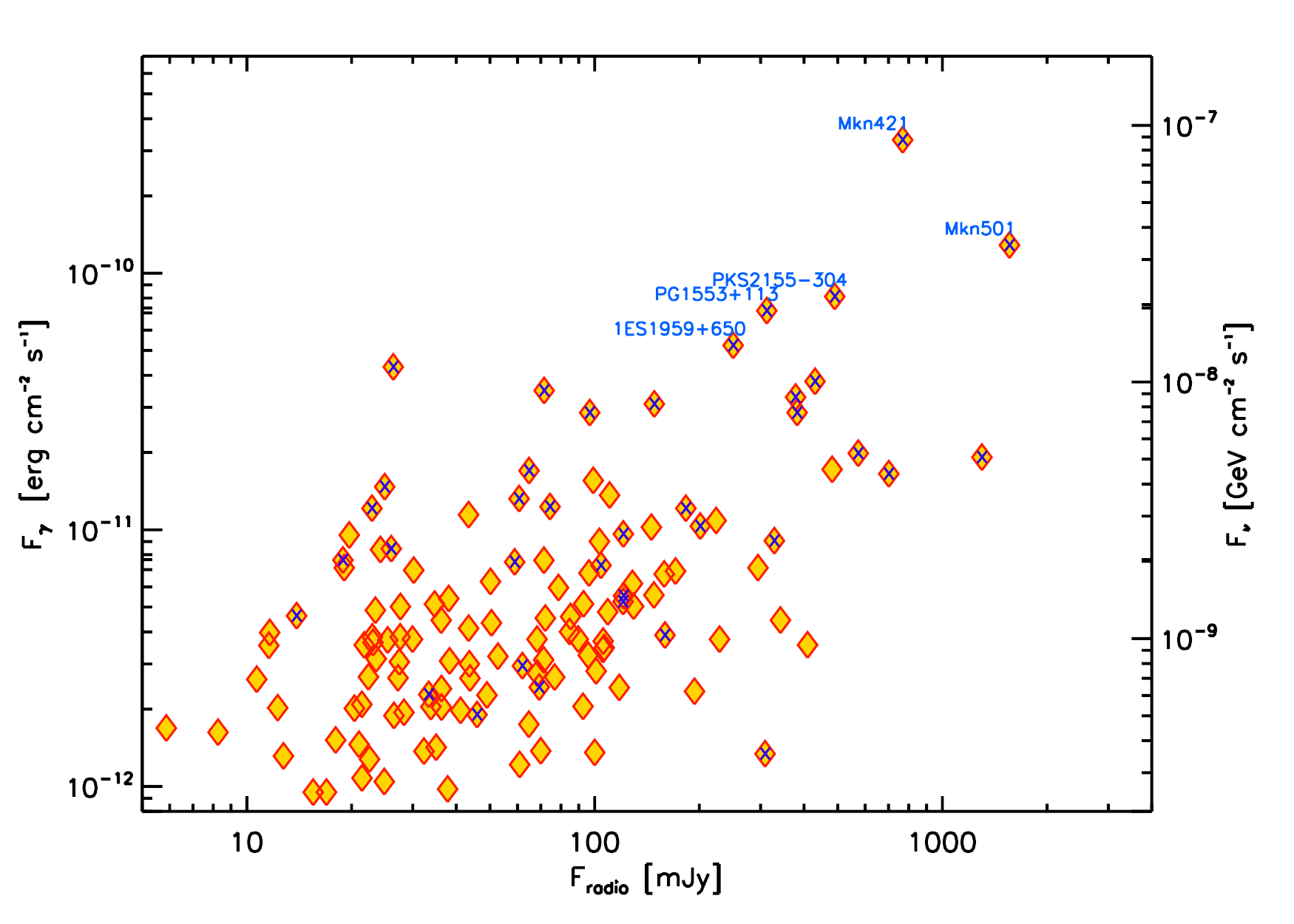}
\caption{Radio flux versus the high-energy $\gamma$-ray flux ($E>50$ GeV) of the 132 HBL belonging to the 2FHL catalogue. Blue crosses indicate sources detected in the TeV band. The vertical axis on the right report the muon neutrino flux (in the 0.1-10 PeV band) predicted with the scaling discussed in the text. We also show the name of the brightest sources.}
\label{frfgamma}
\end{figure}

\longtab{
\begin{longtable}{c | lc | cc | cc}
\caption{\label{TableKM3} Expected 0.1-10 PeV flux (in units of $10^{-8}$ GeV cm$^{-2}$ s$^{-1}$) and detection rate of muon neutrino $R_{\nu}$ (yr$^{-1}$) for the brightest 2FHL BL Lacs with  KM3NeT with different thresholds on the zenith angle (horizon and $+10^{\circ}$). We also report the fraction of the observational time for which each source is below the threshold. The numbers identify the sources in the sky map, Fig. \ref{fig:map}.}\\ 
\hline
\hline
 &  Name & $F_\nu$ & $R_{\nu}$ & Visibility  & $R_{\nu}$ & Visibility   \\
  & & & & at horizon & & at $10^\circ$  \\
  \hline
\hline
\hline
\hline
\hline
 1 & Mkn421 				& 8.77	&  4.59 &       0.30 &        5.80 &       0.39		\\
 2 & PKS2155-304 			& 2.15	&  2.23 &       0.60 &        2.53 &       0.69		\\
 3 & Mkn501 				& 3.41	&  1.65 &       0.28 &        2.26 &       0.39		\\
 4 & PG1553+113 			& 1.89	&  1.42 &       0.44 &        1.66 &       0.51		\\
 5  &PKS0447-439 			& 0.76	&  0.87 &       0.67 &        1.02 &       0.79		\\
 6  &PKS1424+240 			& 1.00	&  0.67 &       0.39 &        0.79 &       0.46		\\
 7 & PKS2005-489 			& 0.51	&  0.63 &       0.72 &        0.75 &       0.86		\\
 8 & TXS0518+211 			& 0.87	&  0.59 &       0.39 &        0.72 &       0.48		\\
 9 & PG1218+304 			& 0.92	&  0.55 &       0.34 &        0.69 &       0.44		\\
 10 &1ES0647+250 			& 0.75	&  0.47 &       0.36 &        0.60 &       0.46		\\
 11 &3C66A 				& 0.87	&  0.38 &       0.25 &        0.54 &       0.36		\\
 12 &1RXSJ054357.3-55320 	& 0.30	&  0.40 &       0.78 &        0.52 &        1.00	\\
 13 &PKS0301-243 			& 0.43	&  0.44 &       0.59 &        0.49 &       0.66		\\
 14 &1H1914-194 			& 0.45	&  0.44 &       0.57 &        0.49 &       0.63		\\
 15$^a$ &1H1013+498		& 0.87	&	- &	-	&        0.48 &       0.32		\\
 15$^b$ &1RXSJ194246.3+10333 &  0.41	&  0.32 &       0.45 &	 -	&	-		\\
 16 &PKS1440-389 			& 0.36	&  0.41 &       0.66 &        0.47 &       0.76		\\
 17 &1ES0347-121 			& 0.39	&  0.35 &       0.53 &        0.40 &       0.60		\\
 18 & 1ES1215+303 			& 0.52	&  0.31 &       0.34 &        0.39 &       0.44		\\
 19 &1RXSJ101015.9-31190 	& 0.32	&  0.34 &       0.60 &        0.39 &       0.69	\\
 20 &RXJ0648.7+1516 		& 0.45	&  0.33 &       0.42 &        0.38 &       0.49	\\
\end{longtable}
}

\section{Application and results}

Padovani et al. (2016) found a significant probability of association between the positions of the HBL belonging to the 2FHL catalogue\footnote{defined as BL Lac with synchrotron peak frequency larger than $10^{15}$ Hz.}  with flux $F(>50 \, {\rm GeV})\gtrsim 2\times 10^{-11}$ ph cm$^{-2}$ s$^{-1}$ and a selected sample of neutrinos \footnote{including both HESE (four years) and through-going $\nu_{\mu}$ (two years).} detected by IceCube above 60 TeV. For illustration, in Fig. \ref{frfgamma} we report the radio flux versus the $\gamma$-ray energy flux (integrated over the 50-2000 GeV band using the spectral parameters of the 2FHL) for the 132 HBL of the 2FHL (for the selection we used the phenomenological estimate of the synchrotron peak frequency provided in the  3rd Catalog of AGN Detected by the {\it Fermi}-LAT, 3LAC in brief).

Our aim is to provide the neutrino counts expected from each 2FHL BL Lacs,  in view of the possible identification of the extragalactic neutrino sources based on a positional correlation with detected neutrinos. For this reason, it is justified to specialize our treatment and focus it on the through-going muon neutrinos, $\nu_{\mu}$. Indeed, muons leave well-defined and long tracks easier to reconstruct, determining the best ($<1^{\circ} $) angular resolution, suitable to look for possible associations with point-like sources as HBL. In the case of IceCube this implies to focus on the component coming from the northern hemisphere. On the contrary, KM3NeT will be sensitive to through-going muon originating mainly from neutrinos coming from the southern hemisphere. 

The IceCube collaboration published both the spectrum of the so-called high-energy starting events (HESE), dominated by cascade-like events triggered within the detector volume by neutrinos from the the southern sky (Aartsen et al. 2015a) and that derived analyzing only the (high-energy, $E\gtrsim 100$ TeV) muon-like northern events (Aartsen et al. 2015b, R{\"a}del \& Schoenen 2015). The derived spectral parameters are in tension, with the through-going muon signal providing a spectrum harder ($\Gamma=1.91\pm 0.20$ using events with $E>170$ TeV, R{\"a}del \& Schoenen 2015)  than that ($\Gamma=2.50\pm 009$, Aartsen et al. 2015a) derived from HESE data with an extension to low energy ($E>60$ TeV). Interestingly, if only high-energy HESE events ($E>100$ TeV) are selected, the tension reduces. This could be considered as an hint for hardening of the spectrum at high-energy, possibly related to two (galactic and extragalactic) spectral components (see the detailed discussion in Palladino \& Vissani 2016). Moreover, if extragalactic, the high  neutrino flux below 60 TeV would imply an accompanying gamma-ray flux exceed the high-energy extragalactic $\gamma$-ray background (e.g. Murase et al. 2016). It is also worth adding that it is unlikely that the neutrino spectrum predicted by the structured jet model extends below 100 TeV, since this would imply a  quite large cosmic-ray power (Paper I). Summing up, there are hints supporting the scenario in which lower energy neutrinos ($\lesssim 100$ TeV)  derive  from another population -- possibly galactic -- of sources. All this justifies the use in the following of the spectral parameters obtained with the through-going muon analysis (we used the parameters derive by the first 4 years of IceCube, R{\"a}del \& Schoenen 2015).

\begin{figure}
\hspace{-0.45truecm}
 \includegraphics[width=10.cm,height=14cm]{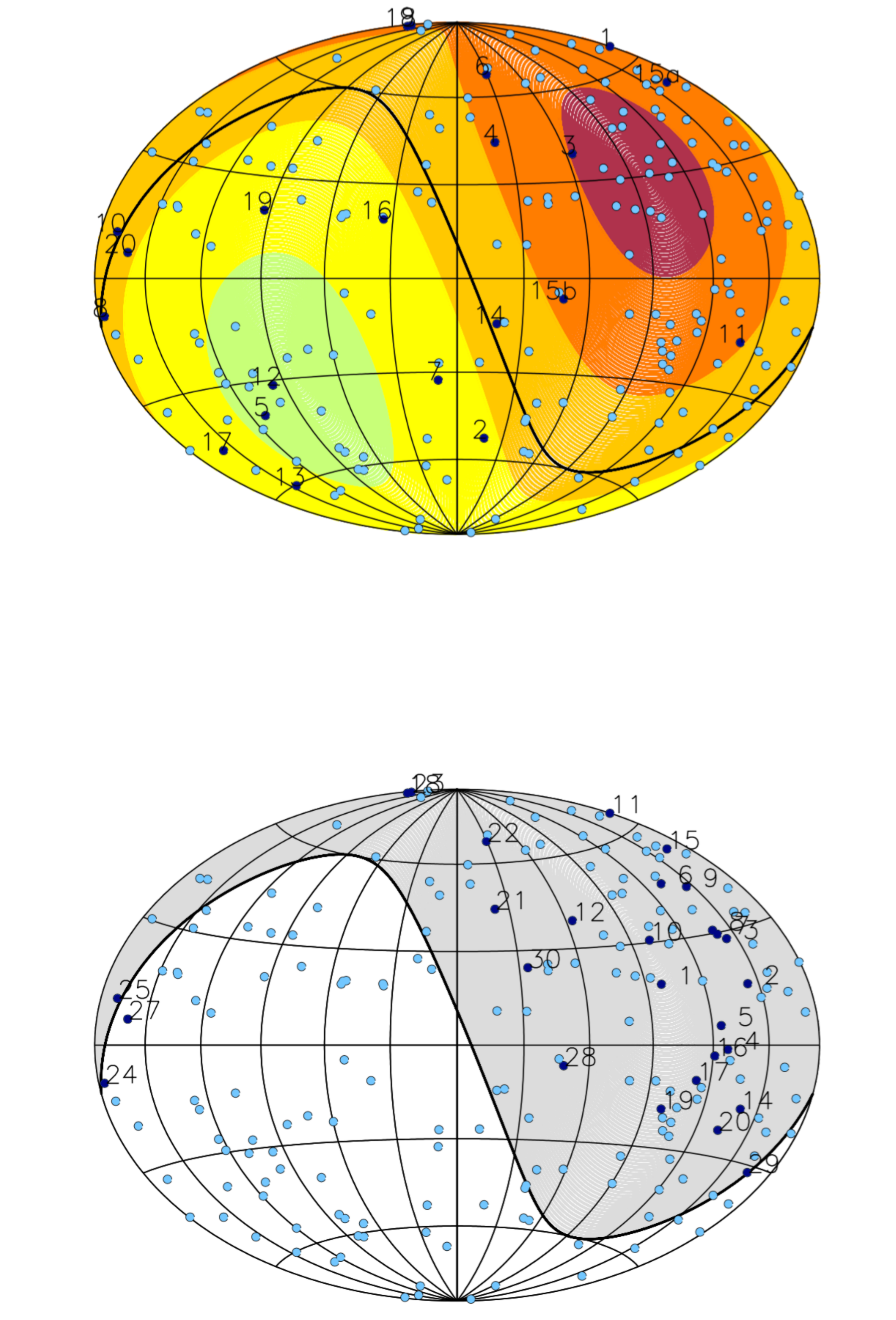}
  \caption{Sky maps in galactic coordinates with the position of 2FHL HBL (light blue and blue) for IceCube (lower panel) and KM3NeT (upper panel). The blue points are the brightest HBL and the associated number is reported in Tables 1 and 2. For the plot of KM3NeT, the different color indicates the different range of object declination-depending visibility; red: $> +60^\circ$ and visibility percentage $< 20\%$, dark-orange: $+25^\circ \div +60^\circ$ and visibility percentage $ 20 \div 45\%$,orange: $-12^\circ \div +25^\circ$ and visibility percentage $ 45 \div60\%$, yellow: $-53^\circ \div -12^\circ$ and visibility percentage $ 60 \div100\%$ and green: $< -53^\circ$ and visibility percentage $100\%$. Lower map: Visibility plot of IceCube. The grey region is $100\%$ of muon neutrino visibility, and the white is near $0\%$.}
  \label{fig:map}
\end{figure}
 
Following the procedure discussed in \S 3, the first step is to derive the constant of proportionality between the $\gamma$-ray and the neutrino flux from Eq. \ref{ftot}.

The total (muon) neutrino flux $F_{\nu,tot}$ is calculated integrating the power law spectrum provided by R{\"a}del \& Schoenen (2015) in the range 100 TeV-10 PeV. The result is $F_{\nu,tot}=4.85 \times 10^{-7}$ GeV cm$^{-2}$ s$^{-1}$. 

The second quantity we need is the total high-energy $\gamma$-ray flux from the neutrino-emitting HBL population, $F_{\gamma,tot}$. An obvious upper bound to this flux is provided by the total (i.e. resolved+unresolved) observed extragalactic high-energy $\gamma$-ray background (Ackermann et al. 2015). Above 50 GeV (the low-energy threshold of the 2FHL) the background intensity is $2.4\times 10^{-9}$ ph cm$^{-2}$ s$^{-1}$ sr$^{-1}$. On the other hand, we calculated that the contribution of the detected HBL of the 2FHL (50-2000 GeV, flux sensitivity limit $\approx 8\times 10^{-12}$ ph cm$^{-2}$ s$^{-1}$) to the background (assuming isotropy) is $7.2\times 10^{-10}$ ph cm$^{-2}$ s$^{-1}$ sr$^{-1}$, corresponding to about $1/3$ of the total background intensity above 50 GeV. Through accurate simulations, Ackermann et al. (2016b) estimated that (resolved and unresolved) point sources with fluxes larger than $10^{-12}$ ph cm$^{-2}$ s$^{-1}$ (the majority of which are assumed to be blazars, but not necessarily all HBL) should account for about $90\%$ of the background. On the other hand, we have also to remind that Padovani et al. (2016) found that the correlation between the IceCube neutrinos and the 2FHL HBL holds only with sources with relatively high-flux ($\gtrsim 1.8\times 10^{-11}$ ph cm$^{-2}$ s$^{-1}$).
Given these uncertainties and in view of the fact that, in any case, the differences involve relatively small factors, in the following we use the value of the flux $F_{\gamma,tot}$ obtained summing the 2FHL BL Lac only, keeping in mind that derived neutrino fluxes should be considered as upper limits since, if also HBL with smaller flux would contribute, the derived neutrino fluxes could be lower by a factor $\approx 3$.  The total energy flux in the 50-2000 GeV band (approximating well the bolometric gamma-ray output, since the high-energy peak is commonly found at 100 GeV) of the 2FHL sources, $F_{\gamma ,tot}$, can be directly performed using the spectral information of the 2FHL, giving $F_{\gamma, tot}= 1.14 \times10^{-6}$ GeV cm$^{-2}$ s$^{-1}$.  Therefore, for the value of the constant we obtain $k_{\nu\gamma}=F_{\nu,tot}/F_{\gamma,tot}= 0.46$.

The vertical axis on the right of Fig. \ref{frfgamma}  reports the neutrino flux for each 2FHL HBL calculated with the scaling above. With these fluxes at hand we can predict the expected count rate for IceCube and KM3NeT. In the following we separately describe the results.

\subsection{IceCube}

IceCube (Achterberg et al. 2006) is a neutrino detector placed at the South Pole. It is the largest operating neutrino detector, encompassing an instrumented cubic kilometer of ice. In the case of $\nu_{\mu}$ -- detected from the upgoing through-going muons -- the effective area of the instrument depends on the declination of the source, since the angle-dependent absorption by the Earth starts to affect the detected flux above $\approx 100$ TeV. The actual effective area in ranges of declinations ($60^{\circ}<\delta<90^{\circ}$, $30^{\circ}<\delta<60^{\circ}$, $0^{\circ}<\delta<30^{\circ}$) are provided by Yacobi et al. (2014). The number of neutrinos expected for the brightest 2FHL sources with an effective exposure of 1 year and divided in ranges of declination are reported in Table \ref{TableIce} and shown in Fig. 2 (lower panel).

Very few sources present a rate exceeding 1 event yr$^{-1}$. Interestingly, among them there are two of the Padovani et al. best candidates, Mkn 421 and PG 1553+113. 1ES 1959+650, from which AMANDA possibly detected three neutrinos during a burst in 2002 (Ackermann et al. 2005) is not expected to be so bright. 
In considering these numbers one should remind that they are upper bounds to the actual values, since, as discussed above (Sect. 4), neutrino fluxes (and count rates) smaller by a factor $\sim$3 are compatible with the $\gamma$-ray background.
Note also that PKS 2155-304, among the brightest 2FHL HBL and thus among the most intense neutrino sources, being a southern object does not enter in our list. 

\subsection{KM3NeT}

KM3NeT (e.g. Margiotta et al. 2014) will be a new undersea neutrino telescope that could detect all-flavour neutrinos. Presently it is under construction in the Mediterranean See. 

The expected effective area as a function of declination, as that used above for IceCube, is not available yet. Therefore we choose to rely on the declination-averaged effective area provided by Adrian-Martinez et al. (2016b). Note that, differently from the case of IceCube, for KM3NeT a given source in the sky is below horizon (and thus the up-going muon technique can be applied) only for a fraction of a year.
Adrian-Martinez et al. (2016b) provide the effective exposure time of sources located at different declinations, that is the fraction of time for which the source is below the horizon and thus data can be obtained. In Table \ref{TableKM3} we thus report the expected neutrino counts for 1 year taking into account the effective exposure of the different sources (also reported in the Table).\\ 

In Fig. \ref{fig:map} we also show a sky map in galactic coordinates reporting the HBL of the 2FHL (light blue points) and in blue the best candidates for KM3NeT (upper panel) and IceCube (lower panel). For KM3NeT the colored areas indicates regions of the sky with different exposures from red (minimum) to green (maximum). In Fig.\ref{frfgamma} we report the calculated neutrino flux as a function of the 2FHL $\gamma$-ray flux for the sources reported in Tables \ref{TableIce}-\ref{TableKM3}.

\section{Discussion}

In this work we have presented a heuristic framework to connect the $\gamma$-ray flux produced through the inverse Compton inside a structured jet of an HBL and the (hypotetical) neutrino flux. 
The scheme is motivated by the findings of Padovani et al. (2016) and is somewhat different from that discussed in Tavecchio et al. (2015). The latter was based on the use of the low-energy gamma-ray band as a proxy for both the cosmic ray power and the density of the photon targets, resulting in a quadratic dependence of the neutrino luminosity on the gamma-ray luminosity. We remark that an important feature of the present scheme, i.e. the dependence on the gamma-ray {\it flux} makes it possible to derive neutrino fluxes also for BL Lacs without a secure redshift measurements. This is quite important, since about 50\% of the HBL of the 2FHL have uncertain $z$.
We also note that, although based on a specific model -- assuming a structured BL Lac jet --, the linear correlation found between $\gamma$-ray and neutrino fluxes have been already suggested in the past for blazars (e.g. Halzen \& Hooper 2005, Neronov \& Ribordy 2009). 

We have derived the expected number of muon neutrino for the HBL of the 2FHL catalogue for both IceCube and KM3NeT. We have provided a list of sources and expected numbers. Our study is focused on the through-going $\nu_\mu$  because of the angular resolution is well defined in the detectors.  Our analysis takes into account the structural differences between the detectors.  We have used the effective area at different declinations for IceCube and the effective area for muon neutrino for all declinations for KM3NeT. A great difference between the two detector is their latitude; IceCube is located at the South Pole, therefore the sources always have the same visibility throughout the year. Different is the case of KM3NeT, that have a range of declinations for which the sources are only partially visible during the year. For this reason we have considered for our calculation the visibility as a function of source declination for the muon-track analysis for tracks below the horizon and up to $10^\circ$ above the horizon, given by KM3NeT collaboration. We have calculated the expected number of neutrino from HBL both for tracks below the horizon and for tracks up to $10^\circ$ above the horizon; the difference between the two values is roughly a factor 1.2. 
A more detailed study will be done when the effective area to the various declinations for KM3NeT will be available.

With our calculation we derive for IceCube fluxes consistent with the observations, predicting that only for few, $\gamma$-ray bright, BL Lacs we expect  a handful of neutrinos detectable in few years of operation.  The majority of the sources, instead, have fluxes implying rates of the order of $\lesssim$0.1 events yr$^{-1}$, for which a clear association is thus problematic.
For KM3NeT, on the other hand, we foresee an appreciable neutrino flux for several sources. We report 20 BL Lacs for which the expected rate is $>0.3$ events yr$^{-1}$. For the brightest sources (Mkn 421, PKS 2155--304, Mkn 501), the event rate would likely be high enough to allow a firm identification.  By construction, our method provides {\it average} fluxes. However, considering a typical flare of HBL, with a factor $\alpha  \gtrsim$10 increase of the gamma (and thus neutrino) flux, lasting for $T_{\rm flare}\sim $1-2 weeks, the neutrino expected to be detected during the flare will be $ N_\nu = \alpha \bigl ( \frac{T_{\rm flare}}{1 \, {\rm yr}} \bigr )\tilde N_{\nu}$ where $\tilde N_{\nu}$ is the annual neutrino counts. This implies that for the handful of sources with an annual count $\tilde N_{\nu} \sim 1$ it would thus be possible to obtain one or more neutrinos concomitantly with the $\gamma$-ray flare. This detection would provide a clear signature that HBL can produce neutrinos.

We would like to remark that, as far as the identification of the sources is concerned, KM3NeT and the proposed upgraded IceCube Gen2 (Aartsen et al. 2014b) are expected to play a quite valuable role. In particular, both are expected to have an improved (sub-degree) angular resolution for through-going muon neutrinos\footnote{The preliminary estimate for Km3NeT is ($<0.2^\circ$) (Adri{\'a}n-Mart{\'{\i}}nez et al.2016b).}, which will greatly help studies of correlation between the direction of the neutrino revealed and an extragalactic (or galactic) source. Moreover, having two instruments covering both hemispheres it will be possible to investigate better possible south-north anisotropies and spectral differences.

The structured jet model that we adopt is based on the assumption that the emission we observe from HBL is (almost) totally produced by leptons through synchrotron and IC mechanisms (although it is applicable to all cases in which one predicts a linear relation between neutrino and $\gamma$-ray fluxes). Protons (or hadrons) are only responsible for the observed neutrino flux. The accompanying UHE $\gamma$-ray photons (from $\pi^0$ decay and emitted by the $e^{\pm}$ pairs from the charged pions decay) are readily reprocessed through electromagnetic cascades, leaving the sources as a low-level MeV-GeV component. This is different from what is instead envisaged in lepto-hadronic models (e.g. Petropoulou et al. 2015, 2016), predicting a luminous, hard MeV-GeV emission. Indeed, observations in the hard-X-ray band by the {\it NuSTAR}) satellite (sensitive up to 80 keV), revealing a steep continuum up to the highest energies, seem to leave small room for this bright hard-X/soft gamma component (e.g. Balokovi{\'c} et al. 2016 for Mkn 421), expected to be have a luminosity not much below that of the observed high-energy peak. Future instruments sensitive in the MeV band will play a key role in clarifying this issue.  In particular, 
the proposed {\it e-ASTROGAM} mission\footnote{\tt http://astrogam.iaps.inaf.it} is foreseen to provide a sensitivity of the order of $\lesssim10^{-12}$ erg cm$^{-2}$ s$^{-1}$  in the band 0.3 MeV--3 GeV, where the bulk of the reprocessed emission is predicted. With such a sensitivity {\it e-ASTROGAM} would be able to detect the reprocessed emission even in the case of moderately bright HBLs.

\begin{acknowledgements}
We are grateful to F. Vissani for useful comments on the manuscript and G. Ghirlanda and G. Ghisellini for discussions. We thank the referee for his/her suggestions that helped us to improve the paper. FT acknowledges  contribution from grant PRIN--INAF--2014. D.G. is supported by a grant from the U.S. Israel Binational Science Foundation. Part of this work is based on archival data and on--line services provided by the ASI Science Data Center.
\end{acknowledgements}

\bibliographystyle{aa} 

\begin{thebibliography}{}

\bibitem[Aartsen et al.(2013)]{2014arXiv1405.5303A} Aartsen, M.~G., Abbasi, R., Abdou, Y.,., et al.\ 2013, Science, 342, 1242856-1

\bibitem[Aartsen et al.(2014)]{2014PhRvL.113j1101A} Aartsen, M.~G., Ackermann, M., Adams, J., et al.\ 2014a, Physical Review Letters, 113, 101101 

\bibitem[IceCube-Gen2 Collaboration et al.(2014)]{2014arXiv1412.5106I} Aartsen, M.~G., et al.\ 2014b, arXiv:1412.5106 

\bibitem[Aartsen et al.(2015)]{2015ApJ...809...98A} Aartsen, M.~G., Abraham, K., Ackermann, M., et al.\ 2015a, ApJ, 809, 98 

\bibitem[Aartsen et al.(2015)]{2015PhRvL.115h1102A} Aartsen, M.~G., Abraham, K., Ackermann, M., et al.\ 2015b, Phys. Rev. Lett., 115, 081102 

\bibitem[Achterberg et al.(2006)]{2006APh....26..155A} Achterberg, A., Ackermann, M., Adams, J., et al.\ 2006, Astropart. Phys., 26, 155 

\bibitem[Ackermann et al.(2005)]{2005ICRC....5....1A} Ackermann, M., Bernardini, E., Hauschildt, T., \& Resconi, E.\ 2005, International Cosmic Ray Conference, 5, 1

\bibitem[Ackermann et al.(2015)]{2015ApJ...799...86A} Ackermann, M., Ajello, M., Albert, A., et al.\ 2015, ApJ, 799, 86 

\bibitem[Ackermann et al.(2016)]{2016ApJS..222....5A} Ackermann, M., Ajello, M., Atwood, W.~B., et al.\ 2016a, \apjs, 222, 5 

\bibitem[Ackermann et al.(2016)]{2016PhRvL.116o1105A} Ackermann, M., Ajello, M., Albert, A., et al.\ 2016b, Phys. Rev. Lett., 116, 151105 

\bibitem[Adri{\'a}n-Mart{\'{\i}}nez et al.(2016)]{2016ApJ...823...65A} Adri{\'a}n-Mart{\'{\i}}nez, S., Albert, A., Andr{\'e}, M., et al.\ 2016a, \apj, 823, 65 

\bibitem[Adri{\'a}n-Mart{\'{\i}}nez et al.(2016)]{2016JPhG...43h4001A} Adri{\'a}n-Mart{\'{\i}}nez, S., Ageron, M., Aharonian, F., et al.\ 2016b, Journal of Physics G Nuclear Physics, 43, 084001 

\bibitem[Ahlers \& Murase(2014)]{2014PhRvD..90b3010A} Ahlers, M., \& Murase, K.\ 2014, \prd, 90, 023010 

\bibitem[Ahlers \& Halzen(2014)]{2014PhRvD..90d3005A} Ahlers, M., \& Halzen, F.\ 2014, \prd, 90, 043005 

\bibitem[Atoyan \& Dermer(2003)]{2003ApJ...586...79A} Atoyan, A.~M., \& Dermer, C.~D.\ 2003, \apj, 586, 79

\bibitem[Balokovi{\'c} et al.(2016)]{2016ApJ...819..156B} Balokovi{\'c}, M., Paneque, D., Madejski, G., et al.\ 2016, \apj, 819, 156 

\bibitem[B{\"o}ttcher et al.(2013)]{2013ApJ...768...54B} B{\"o}ttcher, M., Reimer, A., Sweeney, K., \& Prakash, A.\ 2013, \apj, 768, 54 

\bibitem[Chiaberge et al.(2000)]{2000A&A...358..104C} Chiaberge, M., Celotti, A., Capetti, A., \& Ghisellini, G.\ 2000, \aap, 358, 104 

\bibitem[Essey et al.(2010)]{2010PhRvL.104n1102E} Essey, W., Kalashev, O.~E., Kusenko, A., \& Beacom, J.~F.\ 2010, Physical Review Letters, 104, 141102 

\bibitem[Fossati et al.(1998)]{1998BAAS...30..768F} Fossati, G., Maraschi, L., Ghisellini, G., \& Celotti, A.\ 1998, \baas, 30, 130.03 

\bibitem[Ghisellini et al.(1998)]{1998MNRAS.301..451G} Ghisellini, G., Celotti, A., Fossati, G., Maraschi, L., \& Comastri, A.\ 1998, \mnras, 301, 451 

\bibitem[Ghisellini et al.(2005)]{2005A&A...432..401G} Ghisellini, G., Tavecchio, F., \& Chiaberge, M.\ 2005, \aap, 432, 401 

\bibitem[Giroletti et al.(2004)]{2004ApJ...613..752G} Giroletti, M., Giovannini, G., Taylor, G.~B., \& Falomo, R.\ 2004, \apj, 613, 752 

\bibitem[Halzen \& Hooper(2005)]{2005APh....23..537H} Halzen, F., \& Hooper, D.\ 2005, Astroparticle Physics, 23, 537 

\bibitem[Halzen \& Kheirandish(2016)]{2016arXiv160506119H} Halzen, F., \& Kheirandish, A.\ 2016, (arXiv:1605.06119)

\bibitem[Kadler et al.(2016)]{2016arXiv160202012K} Kadler, M., Krau{\ss}, F., Mannheim, K., et al.\ 2016, arXiv:1602.02012 

\bibitem[Kalashev et al.(2013)]{2013PhRvL.111d1103K} Kalashev, O.~E., Kusenko, A., \& Essey, W.\ 2013, Physical Review Letters, 111, 041103

\bibitem[Kalashev et al.(2014)]{2014arXiv1410.8124K} Kalashev, O., Semikoz, D., \& Tkachev, I.\ 2014, arXiv:1410.8124 

\bibitem[Kimura(2014)]{2014cosp...40E1490K} Kimura, S.\ 2014, 40th COSPAR Scientific Assembly, 40,  

\bibitem[Loeb \& Waxman(2006)]{2006JCAP...05..003L} Loeb, A., \& Waxman, E.\ 2006, \jcap, 5, 003 

\bibitem[Mannheim(1995)]{1995APh.....3..295M} Mannheim, K.\ 1995, Astroparticle Physics, 3, 295 

\bibitem[Margiotta(2014)]{2014NIMPA.766...83M} Margiotta, A.\ (The KM3NeT Collaboration) 2014, Nuclear Instruments and Methods in Physics Research A, 766, 83 

\bibitem[McKinney(2006)]{2006MNRAS.368.1561M} McKinney, J.~C.\ 2006, \mnras, 368, 1561 

\bibitem[Meyer et al.(2011)]{2011ApJ...740...98M} Meyer, E.~T., Fossati, G., Georganopoulos, M., \& Lister, M.~L.\ 2011, \apj, 740, 98 

\bibitem[Murase \& Beacom(2013)]{2013JCAP...02..028M} Murase, K., \& Beacom, J.~F.\ 2013, \jcap, 2, 028 

\bibitem[Murase et al.(2014)]{2014PhRvD..90b3007M} Murase, K., Inoue, Y., \& Dermer, C.~D.\ 2014, \prd, 90, 023007 

\bibitem[Murase et al.(2016)]{2016PhRvL.116g1101M} Murase, K., Guetta, D., \& Ahlers, M.\ 2016, Physical Review Letters, 116, 071101 

\bibitem[M{\"u}ller et al.(2014)]{2014A&A...569A.115M} M{\"u}ller, C., Kadler, M., Ojha, R., et al.\ 2014, \aap, 569, A115

\bibitem[Nagai(2014)]{2014AIPC.1632...88N} Nagai, D.\ 2014, American Institute of Physics Conference Series, 1632, 88 

\bibitem[Neronov et al.(2015)]{2015A&A...575A..21N} Neronov, A., Semikoz, D., Taylor, A.~M., \& Vovk, I.\ 2015, \aap, 575, A21 

\bibitem[Neronov \& Ribordy(2009)]{2009PhRvD..80h3008N} Neronov, A., \& Ribordy, M.\ 2009, Phys. Rev. D, 80, 083008

\bibitem[Padovani \& Resconi(2014)]{2014MNRAS.443..474P} Padovani, P., \& Resconi, E.\ 2014, \mnras, 443, 474 

\bibitem[Padovani et al.(2016)]{2016MNRAS.457.3582P} Padovani, P., Resconi, E., Giommi, P., Arsioli, B., \& Chang, Y.~L.\ 2016, \mnras, 457, 3582 

\bibitem[Palladino \& Vissani(2016)]{2016arXiv160106678P} Palladino, A., \& Vissani, F.\ 2016, ApJ, submitted (arXiv:1601.06678) 

\bibitem[Petropoulou et al.(2014)]{2014MNRAS.445..570P} Petropoulou, M., Giannios, D., \& Dimitrakoudis, S.\ 2014, \mnras, 445, 570 

\bibitem[Petropoulou et al.(2016)]{2016APh....80..115P} Petropoulou, M., Coenders, S., \& Dimitrakoudis, S.\ 2016, Astroparticle Physics, 80, 115 

\bibitem[Petropoulou et al.(2015)]{2015MNRAS.448.2412P} Petropoulou, M., Dimitrakoudis, S., Padovani, P., Mastichiadis, A., \& Resconi, E.\ 2015, MNRAS, 448, 2412 

\bibitem[Piner \& Edwards(2014)]{2014ApJ...797...25P} Piner, B.~G., \& Edwards, P.~G.\ 2014, \apj, 797, 25 

\bibitem[Schoenen \& Raedel(2015)]{2015ATel.7856....1S} R{\"a}del, L. \& Schoenen, S. (for the IceCube Collaboration),\ 2015, proc. of the ICRC 2015 (arXiv:1510.05223)

\bibitem[Rossi et al.(2008)]{2008A&A...488..795R} Rossi, P., Mignone, A., Bodo, G., Massaglia, S., \& Ferrari, A.\ 2008, \aap, 488, 795 

\bibitem[Sbarrato et al.(2014)]{2014MNRAS.445...81S} Sbarrato, T., Padovani, P., \& Ghisellini, G.\ 2014, \mnras, 445, 81 


\bibitem[Tamborra et al.(2014)]{2014JCAP...09..043T} Tamborra, I., Ando, S., \& Murase, K.\ 2014, \jcap, 9, 043 

\bibitem[Tavecchio \& Ghisellini(2008)]{2008MNRAS.385L..98T} Tavecchio, F., \& Ghisellini, G.\ 2008, \mnras, 385, L98 

\bibitem[Tavecchio et al.(2014)]{2014ApJ...793L..18T} Tavecchio, F., Ghisellini, G., \& Guetta, D.\ 2014, \apjl, 793, L18 

\bibitem[Tavecchio \& Ghisellini(2015)]{2015MNRAS.451.1502T} Tavecchio, F., \& Ghisellini, G.\ 2015, \mnras, 451, 1502 

\bibitem[Urry \& Padovani(1995)]{1995PASP..107..803U} Urry, C.~M., \& Padovani, P.\ 1995, \pasp, 107, 803 

\bibitem[Wang et al.(2014)]{2014JCAP...11..028W} Wang, B., Zhao, X., \& Li, Z.\ 2014, \jcap, 11, 028 

\bibitem[Waxman \& Bahcall(1997)]{1997PhRvL..78.2292W} Waxman, E., \& Bahcall, J.\ 1997, Physical Review Letters, 78, 2292 

\bibitem[Yacobi et al.(2014)]{2014ApJ...793...48Y} Yacobi, L., Guetta, D., \& Behar, E.\ 2014, \apj, 793, 48 

\bibitem[Zandanel et al.(2015)]{2015A&A...578A..32Z} Zandanel, F., Tamborra, I., Gabici, S., \& Ando, S.\ 2015, \aap, 578, A32 

\end{thebibliography}

\end{document}